\documentclass[twocolumn,showpacs,preprintnumbers]{revtex4}

\def\be{\begin{equation}}
\def\ee{\end{equation}}
\def\beq{\begin{eqnarray}}
\def\eeq{\end{eqnarray}}
\def\n{\nonumber}
\def\bay{\begin{array}}
\def\eay{\end{array}}

\begin{document}

\preprint{CIRI/02-smw01}
\title{Classical formulation of Cosmic Censorship Hypothesis}

\author{Sanjay M. Wagh}
\affiliation{Central India Research
Institute, Post Box 606, Laxminagar, Nagpur 440 022, India\\
E-mail:ciri@vsnl.com}

\medskip
\date{February 2, 2002}
\medskip

\begin{abstract}
Spacetimes admitting appropriate spatial homothetic Killing
vectors are called spatially homothetic spacetimes. Such
spacetimes conform to the fact that gravity has no length-scale
for matter inhomogeneities. The matter density for such spacetimes
is (spatially) arbitrary and the matter generating the spacetime
admits {\it any\/} equation of state. Spatially homothetic
spacetimes necessarily possess energy-momentum fluxes. We first
discuss spherically symmetric and axially symmetric examples of
such spacetimes that do not form naked singularities for regular
initial data. We then consider the most general spatially
homothetic spacetime and show that the Cosmic Censorship
Hypothesis is {\em equivalent\/} to the statement that gravity has
no length-scale for matter properties.
\\
\centerline{Submitted to: Physical Review D}
\end{abstract}

\maketitle

\newpage \section{Introduction} \label{intro}
The existence of massive stars, star clusters, the galaxies, the
structure hierarchy of the galaxy distribution etc.\ point to the
richness of the phenomenon of gravity, in general. What is
strikingly noticeable in this panorama of the universe is that the
mass and the involved size grow at each of these steps going from
the smaller to the larger objects. This points to the fundamental
nature of the mass and length-scale independence of gravity.
\subsection*{Spatial scale-independence of gravity}
The phenomenon of gravitation does not provide any length-scale or
mass-scale for spatial distributions of matter properties.
Newton's celebrated law of gravitation and its applications in
non-relativistic regime of observations instill sufficient
confidence in this property for us to consider it as one of the
fundamental, observational properties of gravitation. We emphasize
that the scale-independence of Newtonian gravity applies only to
space and not to time. Moreover, Newton's law of gravitation does
not specify any property of matter that it deals with. It applies
irrespective of the form of matter under consideration.

The spatial scale-independence of gravity means that we can
construct a gravitating object of any size and of any mass. It can
be made from any matter. Further, matter within such an object can
be distributed in any desirable manner since gravity does not
provide for the spatial distribution of matter within any
gravitating object. (It is a separate question as to whether every
such object will be stable or not.)

General Relativity is a theory of gravitation. Therefore, if the
spatial scale-independence is any basic property of gravity then,
General Relativity must admit, in general, a spacetime with matter
density as an arbitrary function of {\em each\/} of the three
spatial coordinates. We emphasize that such a spacetime metric and
all other metric forms that are reducible to it under non-singular
coordinate transformations, that is to say, diffeomorphic to it,
are the only solutions of the field equations of General
Relativity that are consistent with gravity not possessing a
length-scale for matter properties.

All other spacetimes that are {\em not\/} diffeomorphic to the
aforementioned spacetime then {\em violate\/} the property that
gravity has no length-scale for matter properties. Further, we
note that spacetimes obtained for matter with some specific
equation of state do {\em not\/} conform with the property of
gravity that it applies to all forms of matter. In short, not all
solutions of the Einstein field equations respect these properties
of the phenomenon of gravitation.

The field equations of General Relativity are based on Einstein's
equivalence principle which is, primarily, the principle of
equality of the {\em inertial\/} and {\it gravitational\/} masses.
The equivalence principle leads to the geometrization of gravity
and, hence, from a variational principle, to the field equations
of General Relativity. However, General Relativity does not
automatically incorporate other basic properties of gravity. This
is evident from the fact that we can always construct a spacetime
violating the spatial scale-independence and equate its Einstein
tensor with the energy-momentum tensor of matter fields to obtain
a solution of the field equations. Therefore, we need to {\em
separately\/} enforce other basic properties of gravity, such as
its spatial scale-independence, on the solutions of the field
equations.

In general, a homothetic Killing vector captures \cite{carrcoley}
the notion of the scale-invariance. A spacetime that conforms to
the spatial scale-invariance, to be called a {\em spatially
homothetic spacetime}, is then required to admit an appropriate
{\em spatial\/} homothetic Killing vector ${\bf X}$ satisfying \be
{\cal L}_{\bf X} g_{ab}\;=\;2\,\Phi\,g_{ab} \ee where $\Phi$ is an
arbitrary constant. We then expect spatially homothetic spacetimes
to possess arbitrary spatial characteristics for matter. This is
also the broadest (Lie) sense of the scale-invariance of the
spacetime leading not only to the reduction of the Einstein field
equations as partial differential equations to ordinary
differential equations but leading also to their separation.

Further, in General Relativity, the newtonian notion of
scale-invariance or self-similarity of a physical problem
\cite{sedov} can be generalized in different possible ways
\cite{carrcoley}. The {\it self-similarity of matter fields\/} is
that the physical quantities transform according to their
respective dimensions. When matter fields exhibit this property of
the scale-invariance, a spatially homothetic spacetime admits, in
addition to the spatial homothetic Killing vectors, other
appropriate homothetic Killing vector(s) and, in this case, we
call the spacetime a {\it source self-similar spacetime}.

In general, a spatially homothetic spacetime is not a source
self-similar spacetime. It must be emphasized that the spatial
homothety is the basic property of gravitation and the
self-similarity of matter fields is an additional restriction on
the spacetime geometry.
\subsection*{Astrophysical considerations}
A {\em physically realistic\/} gravitational collapse problem
imagines matter, with regular initial data, collapsing under its
self-gravity. The resultant compression of matter causes pressure
to build-up in it. Further, matter compression generates heat and
radiation because of either the onset of thermonuclear fusion
reactions or other reasons. The radiation or heat then propagates
through the space. The collapsing matter could stabilize to some
size when its equation of state is such as to provide pressure
support against gravity. If self-gravity dominates, the collapse
continues to a spacetime singularity. The issue of Cosmic
Censorship Hypothesis \cite{penrose1} relates to whether the
singularity is visible to any observer or not, ie, whether it is
naked or not.

Irrespective of what the central object is, matter in the
surroundings will accrete onto it. The accreting matter may,
initially, be dust in the far away regions. However, it gets
compressed as it moves closer to the central object and pressure
must build up in it. In many such situations, heat and radiation
partly escape the system and partly fall onto the central object
together with the accreting matter.

Therefore, any complete spacetime description of the collapse and
accretion processes requires us to properly match different
spacetimes of various such stages, during which the properties of
matter are different from each other, to produce the final
spacetime.  Note that the final spacetime will have to be a
solution of the Einstein field equations. (Note further that the
equation of state at extremely high densities is not known.) In
any case, the final spacetime description of the gravitational
collapse and/or the accretion process must admit a changing
equation of state for collapsing/accreting matter. Further, such a
spacetime must also admit an energy or heat flux during late
collapse or accretion stages. To accomplish this process of
matching different such spacetimes is a herculean, if not
impossible, task.

Hence, another approach to these problems is essential. We could
then demand that a spacetime describing the collapse and/or the
process of accretion in its totality admits {\em any\/} equation
of state and appropriate energy-momentum fluxes. In other words,
the spacetime geometry should be obtainable from considerations
that do not involve the equation of state for the matter in the
spacetime. Furthermore, these considerations should result in a
spacetime admitting energy-momentum fluxes.

We now turn to precisely such considerations in General Relativity
that involve the spatially homothetic spacetimes.

\section{Spherically symmetric spacetime} \label{spherical}
We begin here with a spherically symmetric example of a spatially
homothetic spacetime. We impose \cite{cqg2} a {\em spatial\/}
homothetic Killing vector \be (0, f(r,t), 0, 0)\label{hkvss} \ee
on a general spherically symmetric metric. This {\it uniquely\/}
determines the spherically symmetric metric to that obtained in
\cite{cqg1}, namely
\begin{widetext} \be ds^2=-\,y^2(r)\,dt^2+\gamma^2 \left(
y'\right)^2B^2(t)\,dr^2+y^2(r)\,Y^2(t)\,\left[
d\theta^2+\sin^2{\theta}d\phi^2 \right] \label{ssmetfinal}\ee
\end{widetext} with $f(r,t) = y/(\gamma y')$, a prime
indicating a derivative with respect to $r$ and $\gamma$ being a
constant. (We shall always absorb the temporal function in
$g_{tt}$ by suitable redefinition of the time coordinate. The
coordinates are co-moving.)

The Einstein tensor for (\ref{ssmetfinal}) has components
\begin{widetext} \beq G_{tt}&=& \frac{1}{Y^2}-\frac{1}{\gamma^2B^2} +
\frac{\dot{Y}^2}{Y} + 2\frac{\dot{B}\dot{Y}}{BY}
\\ G_{rr}&=&\gamma^2B^2
\left(\frac{y'}{y}\right)^2
\left[-\,2\frac{\ddot{Y}}{Y}-\frac{\dot{Y}^2}{Y}
+\frac{3}{\gamma^2B^2}- \frac{1}{Y^2}\right]
\\G_{\theta\theta}&=&-\,Y\,\ddot{Y}-Y^2\frac{\ddot{B}}{B}
- Y\,\frac{\dot{Y}\dot{B}}{B}+\frac{Y^2}{\gamma^2B^2}
\\G_{\phi\phi}&=& \sin^2{\theta}\,G_{\theta\theta} \\
G_{tr}&=&2\frac{\dot{B}y'}{By}  \eeq \end{widetext}

Notice that the $t-r$ component of the Einstein tensor is
non-vanishing. Hence, matter in the spacetime could be {\em
imperfect\/} or {\em anisotropic\/} indicating that its
energy-momentum tensor could be \begin{widetext} \beq {}^{\rm
I}T_{ab}&=&(\,p\,+\,\rho\,)\,U_a\,U_b \;+\; p\, g_{ab}
\;+\;q_a\,U_b \;+\; q_b\,U_a \;-\;2\,\eta\,\sigma_{ab} \\ {}^{\rm
A}T_{ab}&=&\rho\, U_a\,U_b \;+\; p_{||}\,n_a\,n_b \;+\;
p_{\bot}\,P_{ab} \eeq
\end{widetext} where $U^a$ is the matter 4-velocity, $q^a$ is the
heat-flux 4-vector relative to $U^a$, $\eta$ is the
shear-viscosity coefficient, $\sigma_{ab}$ is the shear tensor,
$n^a$ is a unit spacelike 4-vector orthogonal to $U^a$, $P_{ab}$
is the projection tensor onto the two-plane orthogonal to $U^a$
and $n^a$, $p_{||}$ denotes pressure parallel to and $p_{\bot}$
denotes pressure perpendicular to $n^a$. Also, $p$ is the
isotropic pressure and $\rho$ is the energy density. Note that the
shear tensor is trace-free. We will represent by $\sigma$ the
shear-scalar that is given by $\sqrt{6}\;\sigma$

Now, the Einstein field equations with imperfect matter yield for
(\ref{ssmetfinal})
\begin{widetext} \beq \rho &=&
\frac{1}{y^2}\left(\frac{\dot{Y}^2}{Y^2} + 2 \frac{\dot{B}}{B}
\frac{\dot{Y}}{Y} + \frac{1}{Y^2} - \frac{1}{\gamma^2 B^2}\right)
\label{sepdens} \\2\,\frac{\ddot{Y}}{Y} \;+\;\frac{\ddot{B}}{B}
&=&\frac{2}{\gamma^2B^2}\;-\; \frac{y^2}{2}\,\left(
\,\rho\,+\,3\,p \right) \label{isopressure}  \\
3\,(2\,\eta)\,\sigma &=&\frac{1}{y^2} \left(\,
\frac{\ddot{B}}{B}\;-\;\frac{\ddot{Y}}{Y}\;+\;\frac{\dot{B}\dot{Y}}{BY}
\;-\;\frac{\dot{Y}^2}{Y^2}\;+\;\frac{2}{\gamma^2B^2}\;-\;\frac{1}{Y^2}\,
\right) \label{ssshear} \\ q &=& - \frac{2 \dot{B}}{y^2 \gamma^2
y' B^3} \label{heatflux} \eeq
\end{widetext}

\noindent where $q^a = (0, q, 0, 0)$ is the radial heat-flux
vector. The radial function $y(r)$ is not determined by the field
equations and the temporal functions $B(t)$ and $Y(t)$ get
determined by the properties of matter such as its equation of
state.

The spacetime of (\ref{ssmetfinal}) conforms with the general
requirements of a physical collapse then. Consequently, the fate
of spherical collapse, ie, whether the collapse results in a black
hole or a naked singularity, can be explored using
(\ref{ssmetfinal}).

The spacetime singularity can result from either $y(r)=0$ for some
$r$ and/or the temporal functions $B(t)$, $Y(t)$ being zero for
some $t=t_o$ in (\ref{ssmetfinal}).
\subsection*{Absence of naked singularities}
The radial null cone equation for (\ref{ssmetfinal}) is \be
\frac{dt}{dr} = \pm\,\gamma \,\frac{1}{y}\left(
\frac{dy}{dr}\right) B(t) \ee and that it is non-singular for any
nowhere-vanishing function $y(r)$.

Hence, there does not exist an out-going null tangent at the
spacetime singularity that results from purely temporal evolution
of these spacetimes if, initially, $y(r)\,\neq\,0$ for all $r$.
From (\ref{sepdens}) it requires the density to be spatially
non-singular always.

However, it is usual  \cite{psj} to enforce on self-similar,
spherically symmetric spacetimes, the form \be \tilde{X}_a =
(T,R,0,0) \label{hkvusual} \ee for the homothetic Killing vector.
But, in \cite{prl2} we showed that, for spherically symmetric
spacetimes, this form (\ref{hkvusual}) of the homothetic Killing
vector is too restrictive and obscures important information about
the properties of such spacetimes, for example, the existence of
naked singularities. This is understandable since the Killing
vector (\ref{hkvss}) corresponds to the {\em simultaneous\/}
scale-invariance of the spacetime in $T$ and $R$ in the sense of
Lie.

We emphasize that, for spherical symmetry, the appropriate form is
(\ref{hkvss}) since it corresponds only to the radial
scale-invariance of the spacetime in the sense of Lie. However,
(\ref{hkvss}) is equivalent to (\ref{hkvusual}) under the
transformation \be R = l(t) \exp\left(\int f^{-1} d
r\right)\label{sstrans1} \ee \be T = k(t) \exp\left(\int f^{-1} d
r\right) \label{sstrans2}\ee Of course, the transformed metric can
always be made diagonal in $R$ and $T$ coordinates. The imposition
of the homothetic Killing vector (\ref{hkvusual}) on a spherically
symmetric spacetime is {\em over-restrictive} and is not demanded
by any {\em basic property of gravitation}. It should be noted
that spacetimes admitting (\ref{hkvusual}) are included in
(\ref{ssmetfinal}) when the transformations (\ref{sstrans1}) and
(\ref{sstrans2}) are non-singular. This simply relates to the
coordinate freedom in General Relativity. Naked singularities of
spacetimes obtained by enforcing only (\ref{hkvusual}) are then
the artefact of the singular transformations (\ref{sstrans1}) and
(\ref{sstrans2}) in some appropriate sense.
\subsection*{Black hole as infinite red-shift surface}
Moreover, a black hole forms \cite{sph-acc} in the spacetime of
(\ref{ssmetfinal}) only as an infinite red-shift surface and not
as a null hyper-surface. This is easily seen by noticing that
$g_{tt}\,=\,-\,y^2$ is non-vanishing at all $r$ for no-where
vanishing radial function $y(r)$. No spatially finite null
hyper-surface then exists with (\ref{ssmetfinal}).

That the infinite red-shift surface forms in (\ref{ssmetfinal})
follows from the vanishing expansion of the radially outgoing null
vector \be \ell^a\partial_a\;=\frac{1}{y}\frac{\partial}{\partial
t} \;+\;\frac{1}{\gamma y' B}\frac{\partial}{\partial r}
\label{rnull} \ee

\noindent of (\ref{ssmetfinal}). The zero-expansion of
(\ref{rnull}) yields a condition only on the temporal metric
functions as \be \frac{\dot{B}}{B}\,+\,2\,\frac{\dot{Y}}{Y}\; =\;
-\frac{3}{\gamma B} \label{ltscon} \ee When this condition is
reached during the gravitational collapse, light and, with it,
matter trapping occurs. It is only at some ``instant'' of the
co-moving time that the curvature becomes strong enough to trap
light and matter. The condition (\ref{ltscon}) determines this
instant of the co-moving time.

The four-velocity of the matter fluid with respect to the
co-moving observer is: \be
U^a\;=\;\left(\,U^t,\,U^r,\,0,\,0\,\right) \ee Defining then the
radial velocity of matter with respect to the co-moving observer
as \be V_r\,\equiv\,U^r/U^t
\ee we then obtain from the metric (\ref{ssmetfinal}):
\beq U^a &=& \frac{1}{y\,\sqrt{\Delta}}\,\left(\,1,\,V_r,\,0,\,0\,\right) \\
\Delta &=&\;1\;-\;\gamma^2\,\left(
\frac{y'}{y}\right)^2\,B^2\,V_r^2 \label{Delta} \eeq

Now, if $d\tau_{\scriptscriptstyle CM}$ is a small time duration
for the co-moving observer and if $d\tau_{\scriptscriptstyle RF}$
is the corresponding time duration for the observer in the rest
frame of matter, then we have \be d\tau_{\scriptscriptstyle
CM}\;=\;\frac{d\tau_{\scriptscriptstyle RF}}{\sqrt{\Delta}}
\label{rshift} \ee Therefore, the co-moving observer waits for an
infinite period of its time to receive a signal from the
rest-frame observer when $\Delta\,=\,0$. Equation (\ref{rshift})
is also the red-shift formula. Clearly, therefore, $\Delta\;=\;0$
is the {\em infinite red-shift surface}.

Of course, the infinite red-shift surface separates the spacetime
of (\ref{ssmetfinal}) into two regions - one that can communicate
to the far away zone and the black hole region that cannot. The
inside and outside of the infinite red-shift surface are then
causally disconnected regions of the spacetime of
(\ref{ssmetfinal}).

We have then the following possibilities \beq
(\Delta\,>\,0)\qquad |\gamma\,(y')\,B\,V_r| \;<\;y \label{bh0} \\
(\Delta\,=\,0)\qquad |\gamma\,(y')\,B\,V_r| \;=\;y \label{bh1} \\
(\Delta\,<\,0)\qquad |\gamma\,(y')\,B\,V_r| \;>\;y \label{bh2}
\eeq Matter with an initial density distribution determined by
$y(r)$ begins to collapse under the condition (\ref{bh0}) with
initial velocity $V_{r, ini}$ and initial heat flux, determined by
$B(t_o)$. The in-fall velocity of matter and heat flux grow as
matter collapse progresses. Matter properties decide whether the
collapse becomes unstoppable or not. Then,  in any unstoppable
collapse, matter reaches the black hole region of (\ref{bh1}) and
(\ref{bh2}) when condition (\ref{ltscon}) is reached.

\section{Axisymmetric spacetime} \label{axisymmeric}
Encouraged by the example  \cite{kill, sfreecollapse, sscollapse}
in spherical symmetry, we then considered \cite{swkgaxi} the
implications of such a requirement of homothety for axially
symmetric spacetimes.

In axial symmetry we have two spatial variables which can be
expected to behave in a homothetic manner, viz. $r$ and $z$. In
other words, we expect the spacetime to admit arbitrary functions
of $r$ and $z$ determining the matter characteristics of axially
symmetric spacetimes. We then consider the axisymmetric metric

\begin{widetext} \be ds^2 = -\,\bar{A}^2(t,r,z) dt^2+
\bar{C}^2(t,r,z) dr^2 + \bar{D}^2(t,r,z) d z^2 +\bar{B}^2(t,r,z)
d\phi^2 \label{asmet1} \ee \end{widetext} and, guided by our
previous considerations \cite{cqg2}, impose the existence of two
independent homothetic Killing vectors of the form
\begin{widetext} \beq H_r &=& (0,f(r),0,0) \label{hkvr} \\
H_z &=& (0,0,g(z),0) \label{hkvz} \eeq \end{widetext} \newpage on
(\ref{asmet1}). The imposition of (\ref{hkvr}) and (\ref{hkvz})
reduces the metric (\ref{asmet1}) {\it uniquely\/} to
\begin{widetext} \beq ds^2 = &-& Z^2(z)\,y^2(r)\, dt^2 \;+\; \gamma_1^2 \,Z^2(z)
\,C^2(t) \,(y')^2\, dr^2\n \\ &+& \gamma_2^2\, D^2(t)\, y^2(r)\,
(\tilde{Z})^2\,dz^2 \;+\; Z^2(z)\,y^2(r)\,B^2(t)\,d\phi^2
\label{aximetfinal} \eeq
\end{widetext}

\noindent where $\gamma$ s are constants, $f(r)=y(r)/(\gamma_1
y')$ and $g(z)=Z(z)/(\gamma_2 \tilde{Z})$, an overhead prime
denotes differentiation with respect to $r$ and an overhead tilde
denotes differentiation with respect to $z$.

The Einstein tensor for (\ref{aximetfinal}) has the following
components
\begin{widetext} \beq G_{tt}&=& -\frac{1}{\gamma_2^2D^2}-\frac{1}{\gamma_1^2C^2} +
\frac{\dot{C}\dot{D}}{CD} + \frac{\dot{B}\dot{D}}{BD}
+\frac{\dot{B}\dot{C}}{BC}\\ G_{rr}&=&\gamma_1^2C^2
\left(\frac{y'}{y}\right)
\left[-\frac{\ddot{D}}{D}-\frac{\ddot{B}}{B} -
\frac{\dot{B}\dot{D}}{BD}+\frac{3}{\gamma_1^2C^2}+
\frac{1}{\gamma_2^2D^2}\right] \\G_{zz}&=&\gamma_2^2D^2
\left(\frac{\tilde{Z}}{Z}\right)\left[-\frac{\ddot{C}}{C}-\frac{\ddot{B}}{B}
- \frac{\dot{B}\dot{C}}{BC}+\frac{3}{\gamma_2^2D^2}+
\frac{1}{\gamma_1^2C^2}\right] \\G_{\phi\phi}&=& B^2\left[
-\frac{\ddot{D}}{D}-\frac{\ddot{C}}{C}
-\frac{\dot{C}\dot{D}} {CD}+\frac{1}{\gamma_2^2D^2} +\frac{1}{\gamma_1^2C^2} \right] \\
G_{tr}&=&2\frac{\dot{C}y'}{Cy} \\
G_{tz}&=&2\frac{\dot{D}\tilde{Z}}{DZ}\\ \n \\
G_{rz}&=&2\frac{\tilde{Z}y'}{Zy} \eeq \end{widetext} where an
overhead dot denotes a time derivative. It is clear from the above
that the spacetime necessarily possesses energy and momentum
fluxes. The matter in the spacetime is {\em imperfect}.

This is interesting in its own right. Any mass-particle of an
axisymmetric body has a Newtonian gravitational force directed
along the line joining it to the origin. This force, which is
unbalanced during the collapse, has generally non-vanishing
components along $r$ and $z$ axes. Hence, a non-static
axisymmetric spacetime of (\ref{aximetfinal}) will necessarily
possess appropriate energy-momentum fluxes!

The coordinates $(t, r, z, \phi)$ are co-moving. The matter
4-velocity, in general, will have all the four components, ie,
$U^a\,=\,\left( U^t, U^r, U^z, U^{\phi} \right)$.

In the case that $U^{\phi}=0$, the spacetime of
(\ref{aximetfinal}) describes any non-rotating, axisymmetric
matter configuration, in particular, a cigar configuration. In the
case that $U^{\phi}\neq 0$, the spacetime of metric
(\ref{aximetfinal}) describes {\em rotating\/} matter
configurations. In other words, it represents the ``internal''
Kerr spacetimes that are also axisymmetric in nature. (It is also
clear that non-static ``internal'' Kerr spacetimes cannot admit
any perfect fluid matter since axisymmetry requires the existence
of appropriate energy-momentum fluxes in such spacetimes as is
evident from the earlier discussion.)

The spacetime (\ref{aximetfinal}) has a singularity when either
$C(t)=0$ or $D(t)=0$ for some $t$ or when $y(r)=0$ for some $r$
and/or $Z(z)=0$ for some $z$.

Moreover, from (\ref{aximetfinal}), the $r$ and $z$ null cone
equations are \beq \frac{dt}{dr} &=& \pm\,\gamma_1 \frac{y'}{y}
C(t) \\\frac{dt}{dz} &=& \pm\,\gamma_2 \frac{\tilde{Z}}{Z} D(t)
\eeq and these are non-singular for nowhere-vanishing functions
$y(r)$ and $Z(z)$. Hence, there does not exist an out-going null
tangent at the spacetime singularity when $y(r) \neq 0$ and $Z(z)
\neq 0$. Hence, the singularities of these axisymmetric spacetimes
are {\em not\/} naked with these restrictions on the spatial
functions.

We note that the nowhere-vanishing of $y(r)$ and $Z(z)$ means that
the density is initially non-singular. Moreover, it is also clear
that the spacetime of (\ref{aximetfinal}) will allow an arbitrary
density profile in $r$ and $z$ since the field equations do not
determine these spatial functions. Further, it is also seen that
the spacetime of (\ref{aximetfinal}) admits {\em any\/} equation
of state for the matter in the spacetime and that the properties
of matter in the spacetime determine the temporal metric
functions. Hence, on the basis of arguments similar in nature to
the spherically symmetric case, only a black hole as an infinite
red-shift surface forms in the axisymmetric collapse of regular
matter distributions in (\ref{aximetfinal}).

We also note that the energy-momentum tensor of the imperfect
matter in the spacetime can contain contributions from the
presence of electromagnetic fields in the matter. (That is why we
listed only the Einstein tensor above.) The spacetime of
(\ref{aximetfinal}) can then be used to describe the process of
accretion of matter onto a rotating black hole. In this context,
we note that the temporal behavior of the spacetime is all that is
determinable from the properties of matter including those of the
electromagnetic fields in the spacetime.
\section{Most general, spatially homothetic spacetime}
\label{hspgen}
In general, we then demand that the spacetime
admitting no special symmetries, that is no proper Killing
vectors, admits {\em three\/} independent homothetic Killing
vectors corresponding to the three dimensions for which gravity
provides no length-scale for matter inhomogeneities. Such a
metric, from the broadest (Lie) sense, admits three functions
$X(x)$, $Y(y)$, $Z(z)$ of three space variables, conveniently
called here, $x$, $y$, $z$, each being a function of only one
variable.

Based on the above considerations, we then demand that there exist
three independent spatial homothetic Killing vectors \beq  H_1 =
(0, f(x), 0, 0) \label{genhkv1} \\ H_2 = (0, 0, g(y), 0)
\label{genhkv2} \\ H_3 = (0, 0, 0, h(z)) \label{genhkv3} \eeq for
the general spacetime metric \be ds^2\;=\;g_{ab}dx^adx^b \ee with
$g_{ab}$ being functions of the coordinates $(t, x, y, z)$. Then,
the spacetime metric is, {\em uniquely\/}, the following
\begin{widetext} \beq ds^2 = &-&\,X^2(x)\,Y^2(y)\,Z^2(z)\,dt^2
\,+\,\gamma_1^2\left(\,\frac{dX}{dx}\,\right)^2Y^2(y)\,Z^2(z)\,
A^2(t) \,dx^2 \n \\&+&
\gamma_2^2\,X^2(x)\left(\,\frac{dY}{dy}\,\right)^2
Z^2(z)\,B^2(t)\,dy^2\,+\,
\gamma_3^2\,X^2(x)\,Y^2(y)\left(\,\frac{dZ}{dz}\,\right)^2C^2(t)\,dx^2
\label{genhsp} \eeq
\end{widetext}
This is the most general spacetime compatible with gravity not
possessing any length-scale for matter inhomogeneities in its {\em
diagonal\/} form.

The coordinates $(t, x, y, z)$ in which (\ref{genhsp}) is
separable are co-moving. Hence, the matter 4-velocity is
$U^a\,=\,(U^t, U^x, U^y, U^z)$ with all the components
non-vanishing in general. Then, using, for example, the software
{$\scriptstyle{\rm SHEEP}$}, it is easy to see that the Einstein
tensor has appropriate components
\begin{widetext} \beq G_{tt}&=&
-\frac{1}{\gamma_1^2A^2}-\frac{1}{\gamma_2^2B^2}
-\frac{1}{\gamma_3^2C^2} + \frac{\dot{A}\dot{B}}{AB} +
\frac{\dot{A}\dot{C}}{AC} +\frac{\dot{B}\dot{C}}{BC}\\
G_{xx}&=&\gamma_1^2A^2 \left(\frac{L_{,x}}{L}\right)^2
\left[-\frac{\ddot{B}}{B}-\frac{\ddot{C}}{C} -
\frac{\dot{B}\dot{C}}{BC}+\frac{3}{\gamma_1^2A^2}+
\frac{1}{\gamma_2^2B^2} + \frac{1}{\gamma_3^2C^2}\right]
\\G_{yy}&=&\gamma_2^2B^2 \left(\frac{M_{,y}}{M}\right)^2
\left[-\frac{\ddot{A}}{A}-\frac{\ddot{C}}{C} -
\frac{\dot{A}\dot{C}}{AC}+\frac{3}{\gamma_2^2B^2}+
\frac{1}{\gamma_1^2A^2} + \frac{1}{\gamma_3^2C^2}\right]
\\G_{zz}&=& \gamma_3^2C^2 \left(\frac{N_{,z}}{N}\right)^2
\left[-\frac{\ddot{A}}{A}-\frac{\ddot{B}}{B} -
\frac{\dot{A}\dot{B}}{AB}+\frac{3}{\gamma_3^2C^2}+
\frac{1}{\gamma_1^2A^2} + \frac{1}{\gamma_2^2B^2}\right] \\
G_{tx}&=&2\frac{\dot{A}L_{,x}}{AL} \\
G_{ty}&=&2\frac{\dot{B}M_{,y}}{BM}\\
G_{tz}&=&2\frac{\dot{C}N_{,z}}{CN} \\
G_{xy}&=&2\frac{L_{,x}M_{,y}}{LM} \\
G_{xz}&=&2\frac{L_{,x}N_{,z}}{LN} \\
G_{yz}&=&2\frac{M_{,y}N_{,z}}{MN} \eeq \end{widetext}
corresponding to expected non-vanishing energy-momentum fluxes.

It is then easy to see that the field equations do not determine
the spatial functions $X(x)$, $Y(y)$, $Z(z)$. Further, the density
is initially non-singular for nowhere-vanishing spatial functions
$X(x)$, $Y(y)$, $Z(z)$. Moreover, the temporal functions $A(t)$,
$B(t)$, $C(t)$ get determined only from the properties of matter
generating the spacetime.

From arguments similar to those considered earlier for spherically
and axially symmetric spacetimes, it then follows that the
spacetime singularity, which results from the vanishing of only
the temporal function(s) $A(t)$, $B(t)$, $C(t)$, is not locally
naked for nowhere-vanishing spatial functions $X(x)$, $Y(y)$,
$Z(z)$. Hence, the most general spatially homothetic spacetime,
(\ref{genhsp}), does not result to a naked singularity in the
gravitational collapse of matter with initially non-singular
properties. A black hole then always results in the gravitational
collapse of matter fields with non-singular spatial properties in
(\ref{genhsp}).

\subsection*{Semi-stable objects} \label{semi-stable}
We note that a collapsing object could stabilize, for some
co-moving time, by the switching on of some forces opposing
gravity. Stable such objects correspond to static spacetimes.
Then, by considering temporal functions of the spatially
homothetic spacetimes, namely, eqs. (\ref{ssmetfinal}),
(\ref{aximetfinal}) and (\ref{genhsp}), appearing in the energy
fluxes to be constants, we could obtain the spacetimes of
stabilized objects with corresponding symmetries. However, these
are everywhere static spacetimes and, hence, not realistic and not
interesting.

When the equation of state of matter in a spatially homothetic,
non-static spacetime approximates to that of the corresponding
static spacetime during the collapse, we may obtain a
semi-stabilized object within these solutions. For such
semi-stable objects, we have, in general, non-vanishing heat
generation in the matter. Such objects may remain ``stable'' for a
long duration of the co-moving time but may, eventually, collapse
due to accretion of matter onto them.

In general, matter collapse may begin as dust but pressure must
build up, nucleosynthesis may commence to produce heat and may
result in a semi-stabilized object like a star. The star may
explode to shed some mass or may collapse under its self-gravity.
A black hole as an infinite red-shift surface but not as a null
hyper-surface forms in the unstoppable collapse and may accrete
matter in its surroundings. The spatially homothetic spacetimes
accommodate these features. Their temporal behavior is determined
by the properties of matter such as its equation of state.

\subsection*{Machian nature of (\ref{genhsp})}
Mach's principle is the hypothesis of the relativity of inertia.
In a machian theory, the inertia of a body gets determined by the
presence of all other bodies in the universe.

Mach's principle states that the inertia of a particle of matter
is the result of its interaction with all other particles in the
universe. Consequently, there must be energy density of matter
``everywhere'' in a machian universe. This can be interpreted to
mean that we can assemble ``masses'' to produce another ``mass''
and that the process of this building up of mass cannot be
terminated in space. This is, then, recognized as the principle of
the mass-scale and/or spatial scale invariance of the theory of
gravity. Therefore, {\em the spatial scale-independence of gravity
is (one of) the direct implications of Mach's hypothesis of the
relativity of inertia}. Then, we emphasize that the spacetime of
(\ref{genhsp}) is also Machian \cite{machian}.

\section{Discussion} \label{conclude}

Gravity does not provide any length-scale for matter properties.
This requirement, through spatially homothetic spacetime of
(\ref{genhsp}), is then sufficient to ensure that the spacetime
singularities are not visible to any observers in gravitational
collapse of matter with initially non-singular spatial properties.
Moreover, a spatially homothetic spacetime admits any equation of
state for the matter generating it and, hence, such spacetimes
satisfy the general requirements of astrophysical nature needed to
be imposed on the gravitational collapse problem.

\subsection*{Relation with previous results}
We note that some indications already existed in the literature
that point to some of the results or conclusions obtained here.
For example, a result \cite{mcintosh} belonging to this class is
that a perfect fluid spacetime cannot admit a non-trivial
homothetic Killing vector which is orthogonal to the fluid
4-velocity unless $p=\rho$. The spacetime of (\ref{ssmetfinal})
admits (\ref{hkvss}) - a non-trivial, spatial homothetic Killing
vector orthogonal to the fluid 4-velocity. Then, from
(\ref{sepdens}) and (\ref{isopressure}), it follows that the
equation of state for the matter when the time derivatives vanish
is \be
p\,=\,\frac{1}{y^2}\left(\frac{4}{\gamma^2B^2}\,-\,\frac{2}{Y^2}
\right)\;+\;\rho \label{stateqstate} \ee where $B$, $Y$ and
$\gamma$ are constants. The equation of state for the matter in
(\ref{ssmetfinal}) is (\ref{stateqstate}) when the spacetime is
static, in general. It is {\em uniquely\/} $p=\rho$ since the
constants can be chosen appropriately.

Another result \cite{mcintosh} is that a non-flat {\em vacuum\/}
spacetime can only admit a non-trivial homothetic Killing vector
if that vector is neither null nor hyper-surface orthogonal. We
interpret this result to mean that a vacuum spacetime can admit
spatial homothetic Killing vectors.

There also are the following exceptional situations in which we
need not demand the existence of spatial homothetic Killing
vectors:
\begin{itemize} \item The first one being the
Friedmann-Lemaitre-Robertson-Walker (FLRW) solution. This
spacetime corresponds to the homogeneous and isotropic matter
distribution. Note that it admits only perfect fluid matter with
any equation of state and that the equation of state determines
its temporal evolution. This is also the degenerate metric limit
of the general spatially homothetic spacetime (\ref{genhsp}). This
spacetime need not admit any spatial homothetic Killing vectors.
\item The second one is the case of vacuum spacetimes. The vacuum
spacetimes are not required to admit any spatial homothetic
Killing vectors. When there is no matter in the spacetime, we do
not impose any principle related to matter!
\end{itemize}
These exceptions arise primarily because the spacetime is either
vacuum or has homogeneous and isotropic distribution of matter. In
either situation, there is then no necessity for invoking the
principle of no-length-scale for matter properties since it is
implicitly satisfied by these spacetimes. Moreover, the FLRW
spacetime is also the only non-static, perfect-fluid solution that
is compatible with gravity not possessing any length-scale for
matter inhomogeneities.

Further, the spacetimes admitting homothetic Killing vectors of
the form \be (T, \bar{x}, \bar{y}, \bar{z}) \label{usualgenhkv}
\ee or, combinations thereof, are contained with (\ref{genhsp})
provided the transformations of (\ref{genhkv1}) - (\ref{genhkv3})
leading to (\ref{usualgenhkv}) are non-singular. (In \cite{prl2},
we provided the example of this type for spherically symmetric
spacetimes. This has also been considered earlier in this paper.)
It also generally follows that naked singularities can only arise
in spacetimes for which these transformations are singular.

Hence, there are no spatially regular matter data which result
into naked singularities as end states of gravitational collapse
when spatial scale-invariance is respected as our spherical,
axisymmetric and general examples of spatially homothetic
spacetimes show. Then, all spacetimes reducible to the given
spatially homothetic metrics will not result into naked
singularities for spatially non-singular, regular data of matter
fields.

We must now address the issue of all other solutions of the
Einstein field equations apart from the spatially homothetic
spacetimes. In this connection, we note that solutions of the
field equations obtained for any particular, specific, equation of
state need not be reducible, under non-singular coordinate
transformations, to spatially homothetic spacetimes for the same
equation of state. We emphasize here that such solutions would be
seen to violate the spatial scale-invariance of gravity.

However, solutions obtained for specific equation of state could,
under restrictions, be reducible to the corresponding spatially
homothetic forms, for example, the Vaidya or the Tolman-Bondi
spacetimes. But, we must note that such solutions with specific
equation of state apply only when the equation of state of the
collapsing matter is that of the considered solution. Hence, these
are, under applicable restrictions, only a part of the spatially
homothetic spacetimes that apply to the entire gravitational
collapse problem.

The important point is, however, that the spacetimes that are not
reducible to spatially homothetic spacetimes, namely,
(\ref{ssmetfinal}), (\ref{aximetfinal}) and (\ref{genhsp}),
violate one of the basic properties - the spatial scale-invariance
of gravity. As a result, even the regular initial data for the
matter fields could lead, in such spacetimes, to naked
singularities in some cases and to black holes in some others.
Therefore, if the spatial scale-invariance is any basic property
of gravity then, it is {\it misleading\/} to ask whether the
regular initial data results in a naked singularity or a black
hole as end state of collapse with complete {\em disregard\/} to
this basic property of gravity. Further, if the spatial
scale-invariance is any basic property of gravity then, the
phenomenon of criticality in gravity \cite{choptuik} must also be
reexamined using the spatially homothetic spacetimes.

\subsection*{Importance of spatial scale-independence of gravity}
Some further remarks on the relevance of spatial scale-invariance
and on solutions violating it.

The field equations of General Relativity were arrived at by
demanding only that these reduce to the Newton-Poisson equation in
the weak gravity limit \cite{eg1913, subtle}. But, the {\em field
equations of any theory of gravity should contain the entire weak
gravity physics due to the applicability of the laws of weak
gravity to any form of matter displaying any physical phenomena}.
These equations are only the {\em formal\/} equality of the
appropriate tensor from the geometry and the energy-momentum
tensor of matter. Therefore, the field equations could have been
obtained by imposing the requirement that these reduce to the
single ``equation of the entire weak gravity physics''.

However, there is no ``single'' equation for the ``entire weak
gravity physics" since we include different physical effects in an
ad-hoc manner in the newtonian physics.

But, {\em there can be a ``single'' spacetime containing the
entire weak gravity physics}. Therefore, we need a principle to
identify such a solution of the field equations. In the weak field
limit, the spatial scale-invariance is the freedom of
specification of matter properties through three independent
functions of the three spatial coordinates, in general. To be
precise, we can assemble masses to produce another mass, of any
desired spatial density distribution as well as of any size.
Newtonian law of gravitation permits this even when other physical
phenomena are considered together with that of gravitation.

{\em The spatial scale-invariance is then the principle that could
help us identify spacetimes containing the entire weak gravity
physics.} We have seen in this paper that this is indeed the case
- the spatial scale invariance identifies (\ref{genhsp}) as the
single such spacetime. It has appropriate energy-momentum fluxes,
applicability to any form of matter and, hence, it contains the
entire weak gravity physics.

Clearly, the spatial homothety allows us to distinguish between
solutions that contain the  {\em entire weak gravity physics} and
those {\em that do not}. Spacetimes of the latter kind can only be
of two types - those containing a ``special" part of the weak
gravity physics or ``never" any part of the weak gravity physics.

This is seen as follows. Vacuum spacetimes can never contain any
part of the weak gravity physics. Newton's law of gravity and his
laws of motion have no meaning for vanishing mass. In the same
spirit, solutions for specific equation of state contain
``special" part of the weak gravity physics since these apply to
only considered type of matter. There may be matter solutions
``never" containing any part of the weak gravity physics.

But, {\em why are spacetimes containing ``only a part" and
``never" any part of the weak gravity physics ``physically
not-meaningful"?} We appeal to observations to answer this
question.

Any spatial scale is equivalent with an appropriate mass scale
since the fundamental constants of the theory provide only the
relation leading to the Schwarzschild radius. Then, the spatial
scale-independence of gravity either breaks down at some scale or
it holds at all scales. The break-down of spatial scale invariance
at some scale also implies then the break-down of mass scale. This
means that we cannot assemble masses to form another mass. This
signifies the break-down of the equivalence principle at that
scale. Any such break down has not been observed to 1 part in
$10^{12}$ \cite{cook}.

The equivalence principle and, hence, the spatial scale-invariance
are then fundamental to the theory of gravity. The conclusion that
{\em the spacetimes obeying the spatial scale-invariance are the
only physically meaningful solutions of the field equations\/} is
then inescapable. From our results here, the spacetime of
(\ref{genhsp}) is then the only physically meaningful spacetime.

Many puzzling features may result from the use of ``physically
not-meaningful" spacetimes. As an example, consider the
requirement $g_{tt}\,<\,0$ for all if the spacetime that one
obtains from the principle of equivalence \cite{mitra}. This
requirement implies constraints on the energy-momentum tensor and,
hence, on the forms of matter. But, at the newtonian level, the
law of gravity holds for all forms of matter. This puzzling
feature is a result of the use of physically not-meaningful
spacetimes that have been used in such considerations
\cite{leiter}. Similar puzzling features will also be obtainable
in other theories of gravity if the spatial scale invariance is
not respected. In the same spirit, the existence of naked
singularities is an artefact of the use of ``physically
not-meaningful" spacetimes.

Therefore, the situation with the solutions of the field equations
of General Relativity is understandable only if we realize that
the field equations are based only on the equivalence principle
and do not incorporate the spatial scale-independence of gravity.
In factuality, the newtonian law of gravitation gets replaced by
the single spacetime of (\ref{genhsp}) that contains all of the
weak gravity physics. But, spatial scale-independence needs to be
separately imposed on the field equations to obtain it.

In retrospect, General Relativity replaces a ``single'' law of
weak gravity - Newton's law - with a multiplicity of ``laws of
gravity'' corresponding to many inequivalent spacetimes that are
solutions of the field equations. It is therefore not surprising
that it is only one spacetime, that of (\ref{genhsp}), that alone
truly contains Newton's law, in its entirety, in the weak gravity
limit. In order to identify this `unique' spacetime, we need to
impose the spatial homothety on the field equations because
Newton's law is based on the spatial homothety while the field
equations are more general. Why are spacetimes other than that of
(\ref{genhsp}) are to be considered ``gravitationally
not-meaningful'' or ``physically not-meaningful''? The reason is
then directly related to the fact that some of the results
obtained from such spacetimes will `contradict' the corresponding
results of the weak field theory. Further, the panorama offered by
the universe at small and large spatial scales and also the
experiments related to the testing of the equivalence principle do
not show the break-down of the spatial scale-invariance of gravity
at any scale.

In conclusion, the requirement that General Relativity as a theory
of gravitation does not provide any length-scale for matter
properties results in (\ref{genhsp}). This spatially homothetic
spacetime does not possess a locally or globally naked singularity
for spatially non-singular, regular initial data for matter
fields. Hence, Cosmic Censorship \cite{penrose} is equivalent to
the statement that gravity does not provide any length-scale for
matter properties. We have, in essence, provided also the proof of
this statement here.

\section*{Acknowledgements}

It is a pleasure to thank Kesh Govinder for many stimulating
discussions and useful collaborations. Many extensive and helpful
discussions on various aspects of the Cosmic Censorship Hypothesis
with Ravi Saraykar, Pradeep Muktibodh, Roy Maartens, Pankaj Joshi,
Naresh Dadhich, Sushant Ghosh, Sanjay Jhingan and D. Leiter are
also acknowledged. I also wish to thank D. Leiter for pointing out
some useful references. Further, I am most grateful to Malcolm
MacCallum for providing me the software $\scriptstyle{\rm SHEEP}$
that has been used to perform the calculations presented here.

\end{document}